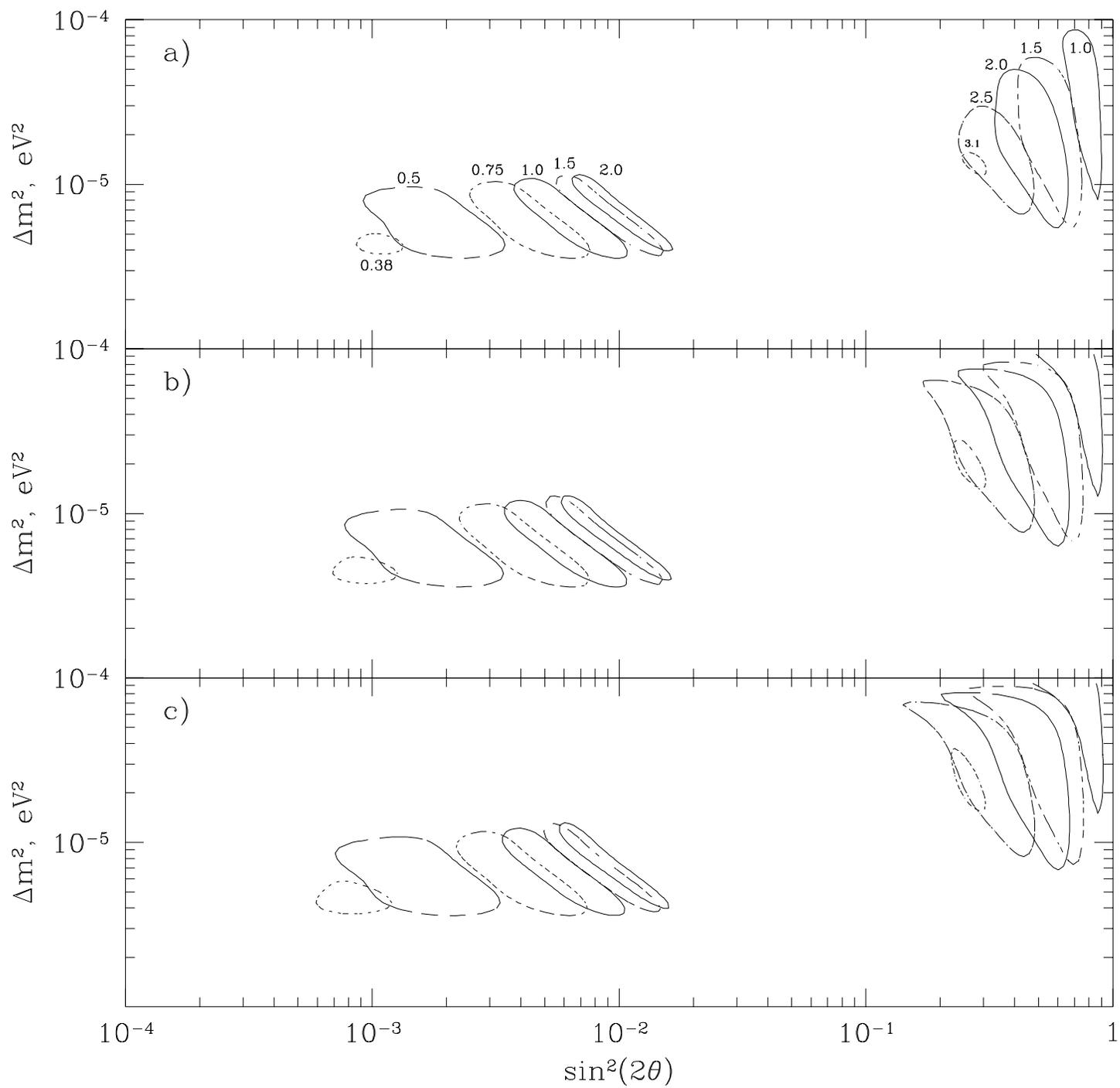

Fig.1

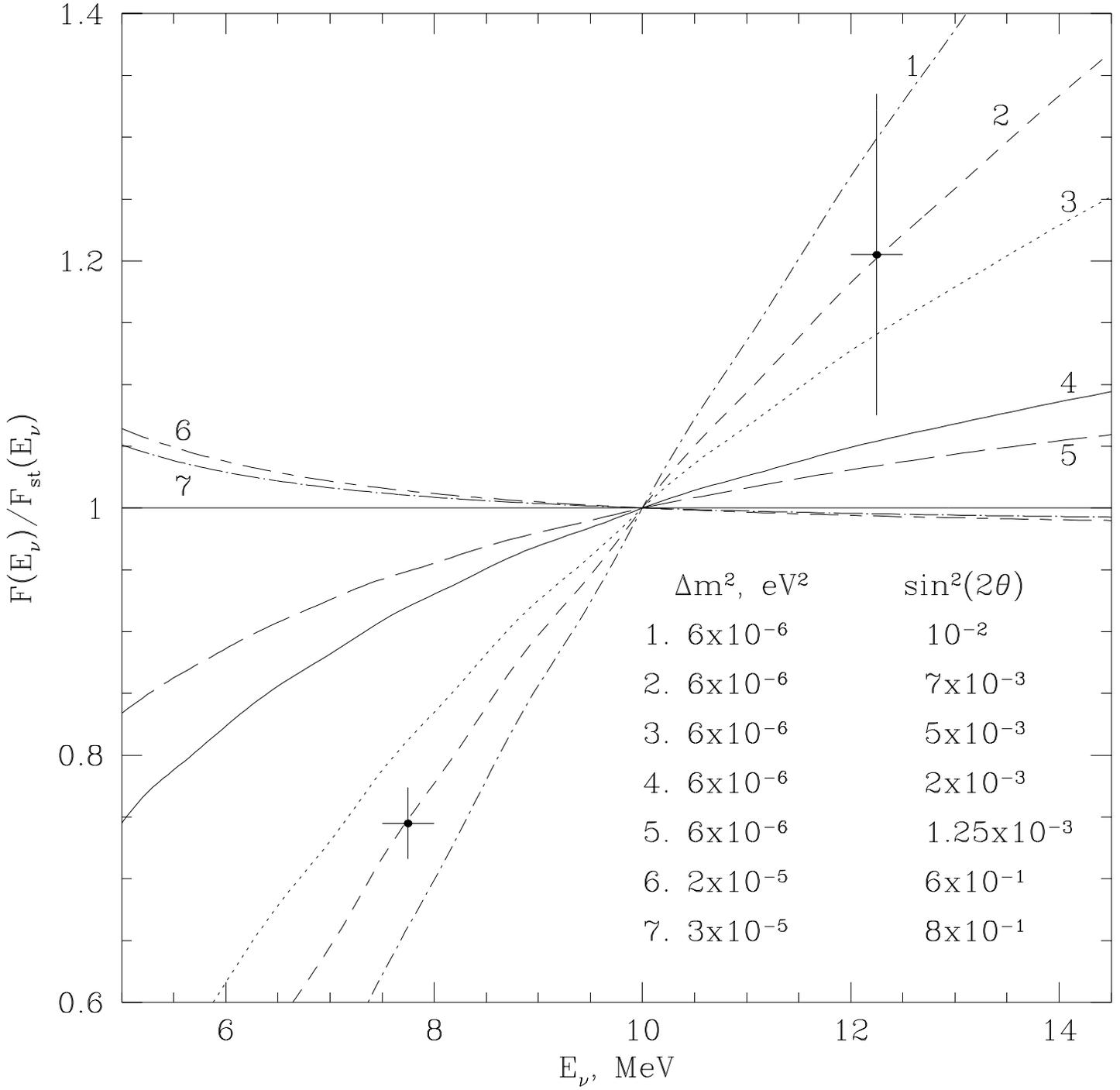

Fig.2

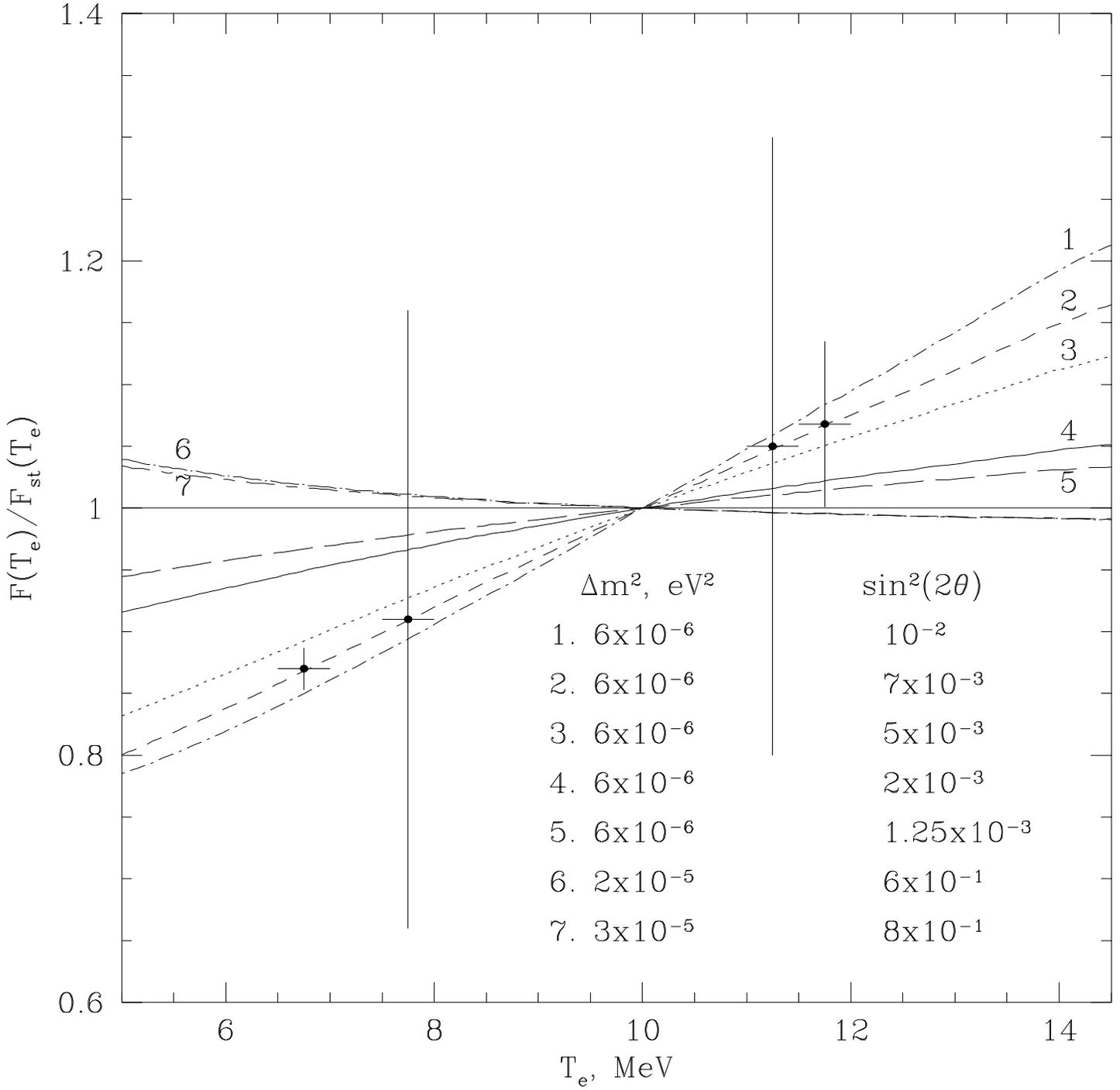

Fig.3

# BORON NEUTRINO FLUX AND THE MSW SOLUTION OF THE SOLAR NEUTRINO PROBLEM

Plamen I. Krastev*

*Institute of Field Physics, Department of Physics and Astronomy,*
*The University of North Carolina at Chapel Hill, Chapel Hill, NC 27599-3255*

Alexei Yu. Smirnov †

*Institute for Nuclear Theory, University of Washington, Henderson Hall HN-12,*
*Seattle, WA 98195,*
*International Center for Theoretical Physics, I-34100 Trieste, Italy*

## Abstract

There are large uncertainties in the predictions of the boron neutrino flux from the Sun which cannot be considered as being of purely statistical origin. We treat the magnitude of this flux, $\Phi_B$, as a parameter to be found from experiment. The properties of the MSW solution to the solar neutrino problem for different values of $\Phi_B$ are studied. Present data give the bounds: $0.38 < \Phi_B/\Phi_B^0 < 3.1$ ($2\sigma$), where $\Phi_B^0 \equiv 5.7 \cdot 10^6$ cm$^{-2}$s$^{-1}$ is the flux in the reference SSM. The variations of the flux in this interval enlarge the allowed region of mixing angles: $\sin^2 2\theta = 8 \cdot 10^{-4} \div 2 \cdot 10^{-2}$ (small mixing solutions) and $\sin^2 2\theta = 0.2 \div 0.85$ (large mixing solution). If the value of the original boron neutrino flux is about that measured by Kamiokande, a consistent description of the data is achieved for $\sin^2 2\theta \sim (0.8 \div 2) \cdot 10^{-3}$ ("very small mixing solution"). The solution is characterized by a strong suppression of the beryllium neutrino line, a weak distortion of the high energy part of the boron neutrino spectrum and a value of the double ratio $(CC/NC)^{exp}/(CC/NC)^{SSM}$ at $E > 5$ MeV close to 1. We comment on the possibility to measure the neutrino parameters and the original boron neutrino flux in future experiments.

---

*On leave from Institute for Nuclear Research and Nuclear Energy, bul. Trakia 72, Sofia 1784, Bulgaria
†On leave from Institute for Nuclear Research, Russian Academy of Sciences, 117312 Moscow, Russia.
e-mail: krastev@physics.unc.edu, smirnov@ictp.trieste.it

# 1  Introduction

Some important aspects of the solar neutrino problem can be formulated in an essentially (solar) model independent way [1] - [5]. However, the implications of the experimental results and, in particular, the appropriate regions of the neutrino parameters in case of neutrino physics solution, strongly depend on the predicted fluxes, the strongest dependence being on the boron neutrino flux, $\Phi_B$. According to the standard solar models [2], [6] –[10] this flux supplies 100%, 70 ÷ 80% and 10% of the Kamiokande, Homestake and gallium signals correspondingly. On the other hand, the predicted value of $\Phi_B$ has rather large uncertainties. They are mainly due to poorly known nuclear cross-sections $\sigma_{17}$, $\sigma_{34}$ at low energies [10], as well as to different astrophysical uncertainties which influence the central temperature of the Sun. In the model [7] these uncertainties are estimated to be at the level of 40%. Some recent experimental [11] and theoretical [12] [13] studies indicate that nuclear cross sections $\sigma_{17}$ and $\sigma_{34}$, extrapolated to solar energies, might be 20 - 40 % below those used in the models [2] - [6]. Furthermore, in [14] it was pointed out, that collective plasma effects have not yet been properly taken into account. A number of small corrections, when added, may result in diminishing of the opacity, and consequently, of the central temperature of the Sun up to 2 - 3%. Thus changes of the cross sections and the implementation of plasma effects may reduce the predicted boron neutrino flux to the one measured by the Kamiokande collaboration. On the other hand, this flux can increase by as much as 20 % in solar models with some mixing and diffusion of elements [15].

In [16] $\Phi_B$ was considered as a free unknown parameter. The $\chi^2$-fit of present data was done in terms of neutrino oscillations parameters $\Delta m^2$, $\sin^2 2\theta$ and the flux $\Phi_B$. In particular, the bound has been obtained $\left(1.43^{+0.65}_{-0.42}\right) \cdot \Phi_B^0$ ($1\sigma$), where $\Phi_B^0 = 5.7 \cdot 10^6$ cm$^{-2}$ s$^{-1}$.

The uncertainties related to the other solar neutrino fluxes are rather small. For the beryllium neutrino flux the predictions from different models have only 10% spread. Other fluxes, like those from $^{13}$N and $^{15}$O decays, although being very uncertain, give small contributions to the signals.

Present uncertainties of the $^8$B–neutrino flux cannot be considered as pure statistical ones. Moreover, these uncertainties are unlikely to get significantly smaller in the near future, at least not before data from new solar neutrino experiments become available. For this reason, we will discuss a solution of the problem without referring to the original (theoretical) value of the boron neutrino flux. This flux will be considered as a *fixed parameter* which should be measured in solar neutrino experiments. The features of the MSW solution to the solar neutrino problem will be studied for different values of $\Phi_B$. We present simple analytical relations which describe the modifications of the solutions when the flux is changed. The range of $\Phi_B$ will be found for which a consistent description of present data exists.

## 2 Boron neutrino flux and solar neutrino signals

We will vary the fluxes with respect to the central values predicted by the "reference model" [6]. The predictions of the reference model will be denoted by a superscript "0".

Let us introduce the parameter $f_B$, so that the original boron neutrino flux equals

$$\Phi_B = f_B \cdot \Phi_B^0, \qquad (1)$$

where $\Phi_B^0 \equiv 5.69 \cdot 10^6$ cm$^{-2}$s$^{-1}$ [6] and $f_B$ is the boron neutrino flux in units of $\Phi_B^0$.

The effect of the resonant flavor conversion, $\nu_e \to \nu_\mu(\nu_\tau)$ [17] is described by the survival probability, $P_B(E)$, where $E$ is the neutrino energy. In case of small mixing solutions the boron neutrino spectrum is on the nonadiabatic edge of the suppression pit. The shape of this edge can be described down to $\sin^2 2\theta \sim 10^{-3}$ by the Landau-Stueckelberg-Zener formula [18] [19]:

$$P_B(E) = P_{LZ} \equiv exp\left(-\frac{E_{na}}{E}\right), \qquad (2)$$

where

$$E_{na} \approx \Delta m^2 l_n \sin^2 2\theta, \qquad (3)$$

where $l_n \equiv \left|\frac{d}{dx}\ln n_e\right|^{-1}$ and $n_e$ is the electron number density. For smaller values of mixing, $\sin^2 2\theta < 10^{-3}$, the survival probability as given by the simple analytical expressions in refs.[18], [20] may differ by as much as 30% from the correct ones. The analytical description of the neutrino transitions found in [21] is free of this shortcoming and can be used in this region. From (2) and (3) one finds the relation between mixing angle and suppression:

$$\sin^2 2\theta \approx \frac{E}{\Delta m^2 l_n}\ln P_B(E), \qquad (4)$$

i.e. the weakening of suppression for fixed $\Delta m^2$ implies a decrease of $\sin^2 2\theta$. From (4) one gets

$$\sin^2 2\theta = \sin^2 2\theta^0 \cdot \left[1 - \frac{\ln f_B}{\ln P_B^0}\right], \qquad (5)$$

where $\theta^0$ and $P_B^0$ correspond to the reference model ($f_B = 1$).

In the region of large mixing solutions the high energy part of the boron neutrino flux is on the bottom of the suppression pit and $P_B$ is practically independent on $E$ : $P_B \approx \sin^2 \theta$.

Let us define the averaged over the neutrino energy survival probabilities for boron (electron) neutrinos in the Homestake, $P_H$, and the Kamiokande, $P_K$, experiments as

$$P_i \approx \frac{\int dE \Phi_B^0 \sigma_i P_B}{\int dE \Phi_B^0 \sigma_i}, \quad (i = H, K). \tag{6}$$

Here $P(E)$ is weighted by the original spectrum and by the relevant cross sections:

$$\sigma_H(E) \equiv \sigma_{Ar}(E), \quad \sigma_K(E) \equiv \int dT_e W(T_e) \cdot \frac{d\sigma_{\nu_e e}(E, T_e)}{dT_e},$$

where $W(T_e)$ takes into account the efficiency of the recoil–electron registration in Kamiokande, as well as the energy resolution of the detector, and $T_e$ is the energy of the recoil electrons. The average probabilities strongly depend on the neutrino parameters. However, their ratio, $P_H/P_K$, is close to 1 and changes rather weakly ($< 10\%$) see Table I. This is related to the fact that the products $\Phi_B^0 \sigma_H$ and $\Phi_B^0 \sigma_K$ depend similarly on energy at $E > 5$ MeV. Both products have peaks with maxima at 10 MeV and 10.5 MeV [22]. The first (Homestake) peak is appears at lower energies, and consequently for the nonadiabatic edge, where P(E) increases with E, one gets $P_H < P_K$. The stronger the distortion of spectrum, the larger the difference. The difference disappears, $P_H/P_K \to 1$, when the distortion becomes weaker $P_B(E) \approx const$. These effects can be seen in Table I: $P_H/P_K$ approaches 1 with diminishing of mixing. In the large mixing domain $P_H/P_K = 1.0$.

In terms of $f_B$ and $P_H$ the $^{37}$Ar production–rate, $Q_{Ar}$, can be written as

$$Q_{Ar} = f_B P_H Q^0_{Ar,B} + Q_{Ar,r}, \tag{7}$$

where $Q^0_{Ar,B} \equiv \int dE \Phi_B^0 \sigma_H = 6.2$ SNU is the contribution of boron neutrinos according to the reference model and $Q_{Ar,r}$ is the real contribution to $^{37}$Ar production–rate from all but the boron neutrino fluxes ("rest" fluxes). The suppression factor for $^{37}$Ar production–rate with respect to the reference value, $Q^0_{Ar,B} \equiv 8.0$ SNU, is

$$R_{Ar} \equiv \frac{Q_{Ar}}{Q^0_{Ar}} = f_B P_H k_B + Q_{Ar,r}/Q^0_{Ar}, \tag{8}$$

here $k_B \equiv Q^0_{Ar,B}/Q^0_{Ar} = 0.775$ is the part of the boron neutrinos in the total signal according to the reference model.

The signal in the Kamiokande experiment can be written as

$$R_{\nu e} = f_B \left[P_K + r(1 - P_K)\right]. \tag{9}$$

Second term in (9) is the contribution of $\nu_\mu e$-scattering, the factor $r \approx 0.16$ is essentially the ratio of the $\nu_\mu e$ and $\nu_e e$ cross–sections. In (9) we have neglected the very small difference in the averaged over cross–sections survival probabilities for $\nu_\mu$ and $\nu_e$.

Let us also introduce the ratio ("double ratio") of suppression factors (8) and (9):

$$R_{H/K} \equiv \frac{R_{Ar}}{R_{\nu e}} = \frac{f_B P_H k_B + Q_{Ar,r}/Q^0_{Ar}}{f_B \left[P_K + r(1 - P_K)\right]}. \tag{10}$$

From the existing experimental data (Table I) one finds:

$$R_{H/K} = 0.625 \pm 0.11 \quad (1\sigma) \tag{11}$$

The $^{71}$Ga production–rate $Q_{Ge}$ depends rather weakly on $f_B$: according to the experiment $Q^0_{Ge,B}/Q^{exp}_{Ge} < 17\%$, where $Q^0_{Ge,B}$ is the contribution to Ge production rate from $^8$B–neutrinos in the reference model.

The relations obtained above allow to control the effects of the $\Phi_B$ changes on the solar neutrino signals. Evidently, the decrease (increase) of the original boron neutrino flux can be compensated by weakening (strengthening) of the suppression due to the resonant conversion. For fixed $\Delta m^2$ this is achieved by diminishing of the mixing angle. However, the variations of $f_B$ cannot be compensated completely by a change of the survival probability. According to eqs.(7) - (10) there are three reasons for this: (i) other fluxes apart from $\Phi_B$ contribute to $Q_{Ar}$, (ii) $\nu_\mu e$- scattering contributes to the Kamiokande signal and (iii) $P_K$ and $P_H$ are different. As a result of these a constraint on $f_B$ and on the mixing angle emerges.

In what follows we will give the estimates of the mixing angle for fixed $\Delta m^2 \approx 6 \cdot 10^{-6}$ eV$^2$. The mass-squared difference, $\Delta m^2$, is restricted essentially by the results from the gallium experiments which imply that the adiabatic edge ($E_a$) of the suppression pit is between the highest energy of the pp-spectrum and the energy of the beryllium line $E^{max}_{pp} < E_a < E_{Be}$. This gives $\Delta m^2 \sim (6 \pm 2) \cdot 10^{-6}$ eV$^2$. The value of $\sin^2 2\theta$ for other values of $\Delta m^2$ can be estimated by using the "diagonal ambiguity": for boron neutrinos the survival probabilities (8) depend essentially on the product $\Delta m^2 \sin^2 2\theta$.

# 3 Data fit for different values of boron neutrino flux. Very small mixing solution

Present data (Table I) fix uniquely the solar neutrino fluxes which give the best fit:

(i). The boron neutrino flux is $\approx 0.41 \cdot \Phi_B^0$ at $E > 7$ MeV. Moreover, the low energy part of the flux (E $\leq$ 7 MeV) should be suppressed stronger.

(ii). The fluxes of the beryllium, pep–, $^{13}$N–, and $^{15}$O–neutrinos are strongly suppressed with respect to the reference model predictions, so that their contributions can be neglected.

(iii). There is little or no suppression of the pp-neutrino flux.

The conditions (i, ii) allow one to reconcile the Kamiokande and the Homestake results, if there is an additional contribution to the Kamiokande signal (but not to the Homestake one) which is equivalent to the flux $\Delta\Phi_B \approx 0.09\ \Phi_B^0$. For the Homestake and Kamiokande experiments then one gets $Q_{Ar} \sim 2.5$ SNU and $R_{\nu e} \approx 0.5$. According to (i) - (iii) the gallium production rate is $Q_{Ge} = Q_{Ge,pp}^0 + 0.41 \cdot Q_{Ge,B}^0 \approx 77$ SNU, where $Q_{Ge,pp}^0 = 71$ SNU and $Q_{Ge,B}^0 = 14$ SNU are the contributions from pp– and $^8$B–neutrinos in the reference model. Any deviations from the so inferred values of the solar neutrino fluxes give a worse fit of the present data.

The resonant flavor conversion of neutrinos $\nu_e \to \nu_\mu(\nu_\tau)$ [17] allows to satisfy the above conditions (i-iii). The best fit of the data in the reference model ($f_B = 1$) is achieved for $Q_{Ar,r} \approx 0$, $P_H \approx P_K \approx 0.12$.

The allowed regions of neutrino parameters for different values $f_B$ are shown in fig.1. They were calculated by $\chi^2$–method according to the data presented in Table I, and correspond to 95% C.L. The original fluxes have been kept fixed (no astrophysical uncertainties for each contour). For the calculation of the survival probabilities we have used the analytical description outlined in [21].

Consider first the small mixing domain. With diminishing $f_B$ the allowed regions shift to smaller mixing. Expression (5) describes the shift for values of $f_B$ not too close to $P_B^0$ ($f_B > 0.5$) and mixing angles $\sin^2 2\theta = (2 \div 5) \cdot 10^{-3}$. In this case all the changes are related to the narrow energy region of the boron (electron) neutrinos only. For smaller or larger angles the allowed regions shrink (fig. 1a), and both lower and upper bounds on $\sin^2 2\theta$, and consequently on $f_B$, exist for which one can get a consistent description of the data.

The shrinking of the allowed regions as well as the existence of bounds are related to the three following effects.

(i) With diminishing $f_B$ and, consequently, mixing angle the contribution of $\nu_\mu$ to the Kamiokande signal decreases. According to (9) one has

$$\Delta R_{\nu e}(\nu_\mu) = f_B r(1 - P_K) \approx f_B r \left(1 - \frac{R_{\nu e}}{f_B}\right), \qquad (12)$$

and the last approximate equality in (12) is true if $P_K$ is not too small and $f_B > R_{\nu e}$. Evidently, $\Delta R_{\nu e}(\nu_\mu) \to 0$, when $f_B \to R_{\nu e} \sim 0.4 \div 0.5$. The contribution drops from 16% at $\sin^2 2\theta = 8 \cdot 10^{-3}$ down to 3% at $\sin^2 2\theta = 2 \cdot 10^{-3}$ and smaller than 2% at $\sin^2 2\theta = 10^{-3}$.

(ii) With decreasing $f_B$ the distortion of the high energy part of the boron neutrino spectrum becomes weaker. The spectrum shifts to the upper part of the nonadiabatic edge. The distortion can be characterized by the ratio of the probabilities $P_1$ and $P_2$ at two different energies $E_1, E_2$. Using eqs.(2) and (3) one finds

$$\frac{P_2}{P_1} = P_1^{\frac{E_1}{E_2} - 1}. \qquad (13)$$

With diminishing $f_B$, and therefore increasing $P_1$, the ratio will approach 1: if $E_2/E_1 = 2$, $P_2/P_1 = 1.82, 1.41, 1.20$ and $1.12$ for $P_1 = 0.3, 0.5, 0.7$ and $0.8$, correspondingly. The change of the distortion of spectra is shown for different values of the neutrino parameters in fig. 2. (The corresponding probability can be restored from eqs.(2),(3)).

With weakening of the distortion of spectrum the ratio $P_H/P_K$ approaches 1 (see Table I).

(iii) When $f_B$ decreases the relative contributions, first of all of the $^{15}$O, pep, $^{13}$N and then of the $^7$Be neutrinos, to the Homestake signal (as well as to the gallium one ) increase. Indeed, for $f_B \sim 1$ all the rest neutrino fluxes are on the bottom of the suppression pit, thus being strongly suppressed. With diminishing $f_B$ the nonadiabatic edge shifts to lower energies and at $\sin^2 2\theta < 2 \cdot 10^{-3}$ it reaches the end point of the spectrum of $^{15}$O neutrinos, $E = 1.73$ MeV. With further diminishing of the mixing angle pep- $^{15}$N- and $^7$Be–neutrinos also turn out to be on the nonadiabatic edge. The contributions from the rest fluxes increase quickly: for $\sin^2 2\theta = 2 \cdot 10^{-3}$ we get $Q_{Ar,r} \sim 0.03$ SNU, for $\sin^2 2\theta = 10^{-3}$: $Q_{Ar,O} + Q_{Ar,N} + Q_{Ar,pep} \sim 0.16$ SNU and, moreover, the contribution of $^7$Be–neutrinos becomes appreciable: $Q_{Ar,Be} \sim 0.14$ SNU. The total contribution $Q_{Ar,r} \sim 0.3$ SNU is larger than the $1\sigma$ error.

For $\sin^2 2\theta > 3 \cdot 10^{-3}$, when all the "rest" fluxes are at the bottom of the pit ($Q_{Ar,r} \approx 0$), one gets from (8) and (10):

$$R_{Ar} = f_B P_H k_B \qquad (14)$$

and
$$R_{H/K} \approx \frac{k_B}{1 + r(1 - P_K)/P_H} \ . \quad (15)$$

With diminishing $\sin^2 2\theta$ the double ratio increases. For relatively small $f_B \sim 0.5$ and $\sin^2 2\theta \sim (2 \div 4) \cdot 10^{-3}$ one can neglect the second term in the denominator of (15) (the effect of $\nu_\mu$). In this case the Kamiokande signal and the double ratio are

$$R_{\nu_e} \approx f_B, \quad R_{H/K} \approx k_B \approx 0.78. \quad (16)$$

The last number should be compared with the experimental value (11). The contribution from the "rest" of the fluxes (see (iii)) further increases the double ratio (16), thus worsening the fit. The relative suppression of the Homestake signal becomes weaker. For this reason the allowed region stops to shift and shrinks at $\sin^2 2\theta = 10^{-3}$. The diminishing of the original fluxes of $^7$Be- and $^{13}$N-, $^{15}$O-neutrinos relaxes these effects, and smaller values of $\sin^2 2\theta$ down to $6 \cdot 10^{-4}$ become allowed (fig.1b,c). Let us stress that for small $\sin^2 2\theta$ the size and the position of the allowed regions are sensitive to fluxes of neutrinos from the CNO–cycle.

The influence of the effects (i) - (iii) on the results of Gallium experiments is rather weak. At $\sin^2 2\theta = 10^{-3}$ the contribution from the $^7$Be neutrinos is $Q_{Ge}^{Be} \approx 4$ SNU and from $^{13}$N, $^{15}$O: $Q_{Ge}^{N,O} \approx 2$ SNU which is smaller than $1\sigma$. Therefore present data from gallium experiments do not give additional bounds on the mixing.

Thus a consistent description of present data is possible even if the predicted boron neutrino flux is about that measured by Kamiokande, i.e. $f_B = 0.4 \div 0.5$. In this case the allowed region of lepton mixing is $\sin^2 2\theta = (0.6 \div 3) \cdot 10^{-3}$ and we will refer to it as to *very small mixing solution*. As follows from the above consideration the solution is characterized by weak distortion of the boron neutrino spectrum at $E > 7$ MeV, and small contribution of $\nu_\mu$ to the Kamiokande experiment.

Consider now the case $f_B > 1$. The suppression of the Homestake signal is described by (14). With increasing $f_B$ (diminishing $P_K$) the contribution of the $\nu_\mu$ to the Kamiokande signal increases and therefore the double ratio (10) diminishes. It is the increase of $R_{\nu_e}$ which gives the upper bound on $f_B$ and mixing. Using (14) and (9) we can express $f_B$ in terms of experimentally observable values and the ratio $P_H/P_K$:

$$f_B = \frac{1}{r}\left[R_{\nu_e} - \frac{(1-r)}{k_B}\frac{P_K}{P_H}R_{Ar}\right] \quad (17)$$

This equation allows to estimate the maximal value of $f_B$ which corresponds to a central value $R_{Ar}^{exp}$ and the largest possible value of $R_{\nu e}$ ($\approx 0.65$): $f_B^{max} \sim 2$. In this case $\sin^2 2\theta \approx 2 \cdot 10^{-2}$.

The best fit of present data with two degrees of freedom: $\chi^2 = 0.09 \div 0.12$ is achieved at $f_B = 1.0$ (i.e. at the reference value). For $f_B = 0.75$ we get $\chi^2_{min} = 0.48 \div 0.57$, for $f_B = 0.5$: $\chi^2_{min} = 1.9 \div 2.2$, and for $f_B = 1.5$: $\chi^2_{min} = 0.78 \div 0.88$. According to fig.1 a $2\sigma$-allowed regions exist for

$$0.38 < f_B < 2.5 \tag{18}$$

which corresponds to the region of mixing angles:

$$\sin^2 2\theta = 6 \cdot 10^{-4} \div 2 \cdot 10^{-2} \tag{19}$$

Lower and upper bounds on flux and angle are determined by the lower and upper bounds on the double ratio correspondingly.

The difference between the results (18), (19) and those obtained in [16] is essentially due to the different value of $Q_{Ar}$ as well as to the different treatment of the boron neutrino flux. In our analysis $f_B$ does not participate in the $\chi^2$ fit.

In the region of large mixing solution the pp-neutrinos are outside of the suppression pit, boron neutrinos are on the bottom of the pit and all other neutrinos are in the intermediate region. Consequently, $P_r > P_H = P_K = \sin^2 \theta$. With increasing boron flux ($f_B > 1$) the probabilities $P_i$ decrease and consequently, the contribution from the "rest" of the neutrino fluxes to $Q_{Ar}$ decreases. On the other hand, the contribution of the $\nu_\mu$ to $R_{\nu e}$ increases and the suppression of the pp-neutrino flux becomes weaker ($1 - \sin^2 2\theta/2$). Thus the suppression picture approaches the one for small mixing and the fit becomes better. Best fit, $\chi^2_{min} = 0.11 \div 0.14$, is achieved at $f_B = 1.8 \div 2.0$. With increasing $f_B$ the allowed region is shifted to smaller mixing angles. Values of $\sin^2 2\theta$ as small as 0.2 become possible. As in the case of small mixing the possible increase of $f_B$ is restricted by the contribution to $R_{\nu e}$ from muon neutrinos, which increases with $f_B$. Maximal allowed values are about $f_B \sim 3$. In contrast with small mixing solution here the distortion of boron neutrino spectrum is very weak.

With diminishing $f_B$ the fit gets worse and the allowed region quickly disappears.

## 4 Measuring the original boron neutrino flux

**1.** The original boron neutrino flux, or equivalently $f_B$, can be measured by a signal which is solely due to neutral currents, if there are no sterile neutrinos in the solar neutrino flux. (The experimental criteria from which the existence of such neutrinos can be inferred have been discussed in [27]). After five years of operation of the SNO detector [28] $f_B$ will be measured with an 1% accuracy by the measurement of the event rate $N_{NC}$ (estimated 1400 ev/year for a boron neutrino flux of $3 \times 10^6$ cm$^{-2}\cdot$ sec$^{-1}$) due to the neutral current disintegration of the deuteron: $f_B \propto N_{NC} \left[ \int dE \sigma_{NC} \Phi_B^0 \right]^{-1}$.

The double ratio

$$R_{CC/NC} \equiv \left(\frac{CC}{NC}\right)^{exp} / \left(\frac{CC}{NC}\right)^{SSM} = P_{SNO} \tag{20}$$

gives immediately the average survival probability in SNO experiment, $P_{SNO}$, provided there is no sterile neutrinos. The parameter $f_B$ is fixed then by the total number of CC–events: $f_B = N_{CC} \left[ P_{SNO} \int dE \sigma_{CC} \Phi_B^0 \right]^{-1}$. Here $N_{CC}$ is the number of the charged current events and $\sigma_{CC}$ is the corresponding cross section. The fluxes measured by NC– and by CC–disintegration of the deuteron in the SNO experiment should approach each other with diminishing $f_B$, i.e. the double ratio will converge to 1. For $f_B = 1, 0.75, 0.5, 0.4$, one gets $R_{NC/CC} = 0.4, 0.6, 0.75, 0.85$ correspondingly.

**2.** The survival probability and, consequently, $f_B$, can be obtained by measuring the distortion of the neutrino energy spectrum. As follows from fig.2 the SNO–detector will be able to distinguish between the distorted spectrum for $\sin^2 2\theta > 6 \cdot 10^{-3}$ and the undistorted one predicted by large and very small mixing solutions. However, it will be difficult to further resolve the spectra corresponding to large mixing, to very small mixing and to astrophysical solutions.

**3.** The recoil–electron spectrum measurements also can fix to some extend the distortion of neutrino spectrum and the neutrino parameters. In fig. 3 the recoil energy spectra are shown for different values of the mixing angle. Also, present Kamiokande sensitivity, as well as the sensitivity of Superkamiokande [29], are depicted. As follows from the figure, the Superkamiokande experiment will be able to distinguish the undistorted spectrum from the distorted one up to $\sin^2 2\theta = 5 \cdot 10^{-3}$. It certainly should be possible to identify small mixing solutions. However, as in case of SNO, it will be difficult to disentangle the large mixing, very small mixing and astrophysical solutions by studying of the high energy part of the solar neutrino spectrum. In this case one can use the information from neutral current measurements as for very small mixing the effects from muon neutrinos are very small in contrast with the large mixing case.

The very small mixing solution can be identified by the strong suppression of the beryl-lium neutrino flux [30].

## 5 Implications

The shift of the allowed region towards smaller mixing angles may have serious implications for particle physics. Suppose the solar neutrinos undergo the conversion $\nu_e \to \nu_\mu$, then for the $\nu_e - \nu_\mu$ mixing, one can write the formula

$$\theta_{e\mu} = \left| \sqrt{\frac{m_e}{m_\mu}} - e^{i\phi} \theta_\nu \right|, \qquad (21)$$

where $m_e$ and $m_\mu$ are the masses of the electron and muon, $\phi$ is a phase and $\theta_\nu$ is the angle related to diagonalization of the neutrino mass matrix. The relation (21) between the angles and the masses is similar to the relation in the quark sector which follows naturally from the Fritzsch ansatz for mass matrices. Such a possibility can be realized in terms of the see-saw mechanism of the neutrino mass generation. According to eq.(21) for very small lepton mixing one needs a strong cancellation of the contributions. The level of the fine tuning can be characterized by the parameter

$$\xi = \frac{\theta_{e\mu}}{\sqrt{m_e/m_\mu}}. \qquad (22)$$

For example, at $\sin^2 2\theta_{e\mu} = 10^{-3}$ one gets $\xi = 10^{-1}$, i.e. the two terms in (21) should be tuned with 10% accuracy. No tuning is needed for $\sin^2 2\theta > 5 \cdot 10^{-3}$.

Note that the values of mixing angles in (19) are still much larger than the mixing between the first and the third generations of the quarks: $\sin^2 2\theta \sim 10^{-5}$. However, the lepton mixing can be easily enhanced by the see-saw mechanism itself with a wide class of right handed neutrino mass matrices (see e.g. [31]). In this case the solar neutrino deficit could be explained by $\nu_e \to \nu_\tau$ conversion. Such a scenario can be realized in the supersymmetric $SO(10)$ with unique scale of symmetry violation.

In the region of large mixing solutions the increase of $f_B$ improves the fit, and moreover, the mixing parameter as small as $\sin^2 2\theta \sim 0.2$, i.e. of the order of the Cabibbo mixing becomes allowed.

# 6  Conclusion

**1.** At present the uncertainties in the predictions of the boron neutrino flux are rather large and cannot be considered as purely statistical. In this connection we have studied the MSW solution of the solar neutrino problem for different but fixed values of the original $^8$B–neutrino flux. Present data (if correct) allow to obtain a bound on this flux, $f_B = 0.38 \div 3.1$ and on the mixing angles: $\sin^2 2\theta = 6 \cdot 10^{-4} \div 2 \cdot 10^{-2}$ in small mixing domain or $\sin^2 2\theta = 0.2 \div 0.9$ in the large mixing domain.

**2.** If the original boron neutrino flux is close to that measured by Kamiokande the data can be described by "very small mixing solution": $\sin^2 2\theta = (0.8 \div 2) \cdot 10^{-3}$ which corresponds to a narrow suppression pit. In this case one expects a weak distortion of the high energy part of $^8$B–neutrino spectrum and a small contribution to the signals at $E > 5$ MeV due to muon neutrinos. For very small mixing angles $\sin^2 2\theta < 2 \cdot 10^{-3}$ the size and the position of the allowed region depend essentially on the fluxes of $^{13}$N–, $^{15}$O– and pep–neutrinos.

**3.** The values of $f_B > 1$ allow for a much better fit of the data in the large mixing domain. The description becomes as good as in the case of a small mixing solution. In particular, one can get strong suppression of the $^7$Be–neutrino flux. The allowed region of the parameters shifts to smaller mixing angles, so that $\sin^2 2\theta = 0.2$ are not excluded.

**4.** The measurement of purely neutral current induced events as well as the double ratio will allow to determine the original flux of boron neutrinos. The study of the neutrino energy spectrum in future experiments will allow also to measure the mixing angle (at fixed $\Delta m^2$) up to $(3 \div 5) \cdot 10^{-3}$, thus identifying the small mixing solution. However, by studying only the spectrum it will be difficult to disentangle the large mixing , very small mixing and astrophysical solutions. In order to distinguish between these solutions, one needs precise measurements of the neutral current effects as well as the measurements of the beryllium neutrino flux.

## Acknowledgments

We are grateful to S. Bilenky, C. Charbonelle, S. Degl'Innocenty, C. Giunti, N. Hata, S. Petcov and W. Haxton for valuable discussions. We thank the Institute for Nuclear Theory at the University of Washington for hospitality. P. Krastev acknowledges the kind hospitality of the Theoretical Physics Division at Fermilab where this work was completed. This work was supported by the U.S. Department of Energy under Grants DE-FG06-90ER40561 and DE-FG05-85ER-40219 and by a grant from the North Carolina Supercomputing Program.

**Table I.**

The averaged survival probabilities for boron neutrinos in the Homestake $P_H$ and the Kamiokande $P_K$ experiments for different values of $\sin^2 2\theta$ and $\Delta m^2 = 6 \cdot 10^{-6}$ eV$^2$.

| $\sin^2 2\theta$ | $P_H$ | $P_K$ | $P_H/P_K$ |
|---|---|---|---|
| $1.25 \cdot 10^{-3}$ | 0.824 | 0.834 | 0.988 |
| $2 \cdot 10^{-3}$ | 0.734 | 0.749 | 0.980 |
| $5 \cdot 10^{-3}$ | 0.465 | 0.487 | 0.955 |
| $7 \cdot 10^{-3}$ | 0.345 | 0.366 | 0.943 |
| $10^{-2}$ | 0.222 | 0.240 | 0.925 |
| $2 \cdot 10^{-2}$ | 0.0564 | 0.0630 | 0.895 |
| $3 \cdot 10^{-2}$ | 0.0201 | 0.0220 | 0.914 |

**Table II.**

The solar neutrino data used in the analysis.

| Experiment | Parameter | Result ($1\sigma$) | Reference |
|---|---|---|---|
| Homestake | $Q_{Ar}$, SNU | $2.55 \pm 0.17(stat) \pm 0.18(syst)$ | [23] |
| Kamiokande I+II | $R_{\nu e} \equiv \frac{\Phi^{exp}}{\Phi^0_B}$ | $0.51 \pm 0.04(stat.) \pm 0.06(syst.)$ | [24] |
| GALLEX I+II | $Q_{Ge}$, SNU | $79 \pm 10(stat) \pm 7(syst)$ | [25] |
| SAGE | $Q_{Ge}$, SNU | $74 \pm 19(stat) \pm 10(syst)$ | [26] |

# Figure Captions

**Fig. 1.** Allowed at 95 % C.L. regions of $\Delta m^2$ and $\sin^2 2\theta$ for different values of the parameter $f_B$ (figures at the curves). In fig.1a only the boron neutrino flux varies by a factor of $f_B$ with respect to the standard one, while the other components of the solar neutrino flux have values as in the standard solar model [6]. In fig.1b the beryllium neutrino flux is reduced by 30 % and in fig.1c, in addition to that, the CNO–neutrino flux is reduced by the same factor.

**Fig. 2.** The distortion of the boron neutrino spectrum due to the MSW effect for different values of the neutrino parameters. The ratio of the distorted spectrum to the undistorted one is normalized to 1 at $E = 10$ MeV. The error–bars shown illustrate the expected sensitivity after five years of operation of SNO (in the reaction $\nu_e + d \to p + p + e^-$) and do not take into account any possible systematic errors.

**Fig. 3.** The distortion due to the MSW effect of the spectrum of recoil–electrons scattered by $^8$B–neutrinos in water Cherenkov detectors for different values of the neutrino paremeters. The ratio of the distorted spectrum to the undistorted one is normalized to 1 at $T_e = 10$ MeV. The error bars shown illustrate the present sensitivity of Kamiokande and the expected sensitivity after five years of operation of the Superkamiokande detector.